\documentclass[twocolumn,aps,prb,showpacs,preprintnumbers,amsmath,amssymb,psfrac]{revtex4}

\usepackage{epsfig}
\usepackage{graphicx}
\usepackage{dcolumn}
\usepackage{bm}

\renewcommand{\dag}{^{\dagger}}

\def\gapp{\lower.35em\hbox{$\stackrel{\textstyle>}{\sim}$}}
\def\lapp{\lower.35em\hbox{$\stackrel{\textstyle<}{\sim}$}}

\begin{document}
%

\title{
Entanglement at the boundary of spin chains near a quantum
critical point and in systems with boundary critical points}
\author{T. Stauber and F. Guinea}
\affiliation{Instituto de Ciencia de Materiales de Madrid, CSIC, Cantoblanco, E-28049 Madrid, Spain.}
\date{\today}
\begin{abstract}
We analyze the entanglement properties of spins (qubits) attached
to the boundary of spin chains near quantum critical points, or to
dissipative environments, near a boundary critical point, such as
Kondo-like systems or the dissipative two level system. In the
first case, we show that the properties of the entanglement are
significantly different from those for bulk spins. The influence
of the proximity to a transition is less marked at the boundary.
In the second case, our results indicate that the entanglement
changes abruptly at the point where coherent quantum oscillations
cease to exist. The phase transition modifies significantly less
the entanglement.
\end{abstract}
%
%
%
%
\maketitle
\section{Introduction}
Quantum phase transitions have attracted intense research
activities on various fields of physics\cite{S99}. Whereas
classical phase transitions are driven by thermal fluctuations,
quantum transitions are induced by a parameter which enhances
quantum fluctuations at zero temperature. For simple models there
is a correspondence between classical and quantum phase
transitions such that the universal behavior of a $D$-dimensional
quantum field theory corresponds to the critical behavior of a
$D+1$-dimensional classical field theory. There are phenomena,
however, which cannot easily be understood in terms of this
correspondence, like the nature of the entanglement of the ground
state wavefunction (see also \cite{BKV04}).
The entanglement properties of the quantum
wavefunction of a device are crucial for determining its
suitability as part of a quantum computer (see, for instance
\cite{GM02}).

Recently, Osterloh et. al.\cite{Oetal02} discussed entanglement
for the translationally invariant transversal Ising model in one
dimension (see also\cite{ON02}). The authors observed that the
derivative of the concurrence with respect to the coupling
constant scales according to the Ising universality class close to
the quantum phase transition. The concurrence is a measure of
entanglement between only two spin-$1/2$ systems\cite{W98}, but
also suitable to characterize entanglement also between
next-nearest neighbors.\cite{ON02} The entanglement in the
transverse Ising and XY models made up of macroscopic (contiguous)
subsystems has been discussed in\cite{Vetal03}, employing the von
Neuman entropy as measure of entanglement. The analysis of the
entanglement properties near a quantum critical point can be
relevant to the analysis of many quantum algorithms, as the
Hamiltonians used to implement them show gapless behavior at some
point in the computation\cite{LO03a,LO03b}.

We analyze here the entanglement properties of two level systems
which are either attached to bulk systems tuned near a quantum
phase transition, or which undergo a boundary phase transition, as
described, for instance, by the dissipative two level
system\cite{Letal87,W99}, or the Kondo model\cite{H97}. These
localized phase transitions are due to the coupling to a gapless
(critical) environment\cite{bqcp}. We will not consider, on the
other hand, the entanglement between qubits as function of their
separation\cite{VPC04,VMC04}, even though we distinguish
between next and nearest-next neighbors.

We consider in the next section the properties of qubits at the
boundary of the Ising model in a transverse field, i.e., the
model studied in\cite{Oetal02,ON02}. We analyze how the
entanglement between the two last qubits varies as function of the
distance to the quantum critical point of the model. We also allow
the values of the couplings at the boundary to vary. In section
III we discuss the entanglement properties of two spins attached
to a critical reservoir, as the coupling between them varies,
inducing a boundary critical point. We give the main conclusions
of our work in section IV.
\section{The Ising model}
We start from the homogeneous Ising model with open boundary
conditions characterized by the parameter $\lambda$. The two spins
at the end are further connected by an additional coupling
parameter $\kappa$. The Hamiltonian is thus given by
\begin{align}
{\cal H}=-\lambda\sum_{i=1}^{N-1}\sigma_i^x\sigma_{i+1}^x-
\sum_{i=1}^{N}\sigma_i^z-\kappa\sigma_1^x\sigma_N^x
\end{align}
where $\sigma_i^{x,z}$ are the $x,z$-components of the Pauli
matrices. 

To solve the model,\cite{LSM61,P70} we first convert all the spin
matrices to spinless fermions. This is done by performing the
well-known Jordan-Wigner transformation
($\{c_i,c_{i'}\dag\}=\delta_{i,i'}$):
\begin{align}
\sigma_i^x&=\exp\left(i\pi\sum_{j=1}^{i-1}c_j\dag c_j\right)(c_i+c_i\dag)\label{Jordan1}\\
\sigma_i^z&=1-2c_i\dag c_i\label{Jordan2}
\end{align}
As usual, we will neglect the term that involves the operator
$\exp(i\pi{\cal N})$, ${\cal N}=\sum_{i=1}^N c_i\dag c_i$ in order
to preserve the bilinearity of the model. An additional Bogoliubov
transformation then yields (up to a constant)
\begin{align}
{\cal H}=\sum_{n=1}^N\omega_n\eta_n\dag\eta_n\quad,
\end{align}
\begin{align}
\label{eta}
\eta_n=\sum_{i=1}^N(g_{n,i}c_i+h_{n,i}c_i\dag)
\end{align}
where the $g_{n,i}$, $h_{n,i}$, and $\omega_n$ are determined
numerically (for the general case). Due to the unitarity of the
Bogoliubov transformation, Eq. (\ref{eta}) is easily inverted to
yield
 \begin{align}
\label{Bogoljubov}
c_i=\sum_{n=1}^N(g_{n,i}\eta_n+h_{n,i}\eta_n\dag)\quad.
\end{align}

\subsection{Concurrence}
We are interested in the reduced density matrix $\rho(i,j)$
represented in the basis of the eigenstates of $\sigma_z$. It is
formally obtained from the ground-state wave function after having
integrated out all spins but the ones at position $i$ and $j$. As
measure of entanglement, we use the concurrence between the two
spins, ${\cal C}(\rho(i,j))$. It is defined as
\begin{align}
\label{Cdefinition}
{\cal C}(\rho(i,j))=\text{max}\{0,\lambda_1-\lambda_2-\lambda_3-\lambda_4\}
\end{align}
where the $\lambda_i$ are the (positive) square roots of the
eigenvalues of $R=\rho\tilde\rho$ in descending order. The spin
flipped density matrix is defined as
$\tilde\rho=\sigma_y\otimes\sigma_y\rho^*\sigma_y\otimes\sigma_y$,
where the complex conjugate $\rho^*$ is again taken in the basis
of eigenstates of $\sigma^z$. It will be instructive to also
consider the ``generalized concurrence''
\begin{align}
\label{C*definition}
{\cal C}^*(\rho(i,j))=\lambda_1-\lambda_2-\lambda_3-\lambda_4\quad.
\end{align}

The reduced density matrix $\rho(i,j)\to\rho$ - from now on we
drop the indices $i$ and $j$ - can be related to correlation
functions. For this, we write the ground-state wave function as
the superposition of the four states
\begin{align}
\notag
|\psi_0\rangle=|\uparrow\uparrow\rangle|\phi_{\uparrow\uparrow}\rangle+
|\uparrow\downarrow\rangle|\phi_{\uparrow\downarrow}\rangle+|\downarrow\uparrow\rangle|
\phi_{\uparrow\downarrow}\rangle+|\downarrow\downarrow\rangle|\phi_{\downarrow\downarrow}\rangle,
\end{align}
where the first ket denotes the $z$-projection of the two spins at
position $i$ and $j$. The matrix element
$\rho_{\uparrow\uparrow,\downarrow\downarrow}=
\langle\phi_{\uparrow\uparrow}|\phi_{\downarrow\downarrow}\rangle$,
e.g., is thus given by
$\rho_{\uparrow\uparrow,\downarrow\downarrow}=\langle\sigma_i^+\sigma_j^+\rangle$,
where $\sigma_i^\pm=(\sigma_i^x\pm\sigma_i^y)/2$.

Due to the invariance of the Hamiltonian under
$\sigma_i^x=-\sigma_i^x$, at least eight components of the reduced
density matrix are zero (for finite $N$). The diagonal entries
read:
\begin{align}
\rho_1\equiv\rho_{\uparrow\uparrow,\uparrow\uparrow}&=(1+
\langle\sigma_i^z\rangle+\langle\sigma_j^z\rangle+\langle\sigma_i^z\sigma_j^z\rangle)/4\\
\rho_2\equiv\rho_{\uparrow\downarrow,\uparrow\downarrow}&=(1-
\langle\sigma_i^z\rangle+\langle\sigma_j^z\rangle-\langle\sigma_i^z\sigma_j^z\rangle)/4\\
\rho_3\equiv\rho_{\downarrow\uparrow,\downarrow\uparrow}
&=(1+\langle\sigma_i^z\rangle-\langle\sigma_j^z\rangle-
\langle\sigma_i^z\sigma_j^z\rangle)/4\\
\rho_4\equiv\rho_{\downarrow\downarrow,\downarrow\downarrow}
&=(1-\langle\sigma_i^z\rangle-\langle\sigma_j^z\rangle+\langle\sigma_i^z\sigma_j^z\rangle)/4
\end{align}
The non-zero off-diagonal entries are
\begin{align}
\rho_+\equiv\rho_{\uparrow\uparrow,\downarrow\downarrow}&=\langle\sigma_i^+\sigma_j^+\rangle\\
\rho_-\equiv\rho_{\uparrow\downarrow,\downarrow\uparrow}&=\langle\sigma_i^+\sigma_j^-\rangle\quad.
\end{align}
The positive square roots of the eigenvalues of $R$ are then given
by $|\sqrt{\rho_1\rho_4}\pm\rho_+|$ and
$|\sqrt{\rho_2\rho_3}\pm\rho_-|$. Due to the semi-definiteness of
the density matrix $\rho$, we can drop the absolute values, i.e.,
$\sqrt{\rho_1\rho_4}\pm\rho_+$ and $\sqrt{\rho_2\rho_3}\pm\rho_-$.

We now define
$I_1\equiv\rho_1\rho_4-\rho_2\rho_3=4(\langle\sigma_i^z\sigma_{j}^z
\rangle-\langle\sigma_i^z\rangle\langle\sigma_j^z\rangle)$ and
$I_2\equiv\rho_+^2
-\rho_-^2=-\langle\sigma_i^x\sigma_{j}^x\rangle\langle\sigma_i^y\sigma_{j}^y\rangle$.
For a homogeneous model, we have $I_1\geq0$ and
$I_2\geq0$.\cite{P70} The largest eigenvalue of Eq.
(\ref{Cdefinition}) is thus given by
$\lambda_1=\sqrt{\rho_1\rho_4}+|\rho_+|$ and the concurrence reads
\begin{align}
{\cal C}^*(i,j)&=2(|\rho_+|-\sqrt{\rho_2\rho_3})\quad.
\end{align}
We note that the above expression also holds for the generalized
boundary conditions. For a homogeneous system, it can be further
simplified to
\begin{align}
{\cal C}^*(i,j)&=({\cal O}_{i,j}-1)/2
\end{align}
where we introduced the total order ${\cal O}_{i,j}\equiv\sum_{\alpha=x,y,z}|
\langle\sigma_i^\alpha\sigma_{j}^\alpha\rangle|$.

\subsection{Numerical Results}
\subsubsection{Open boundary conditions}
\begin{figure}[t]
  \begin{center}
    \includegraphics*[width=3in,angle=0]{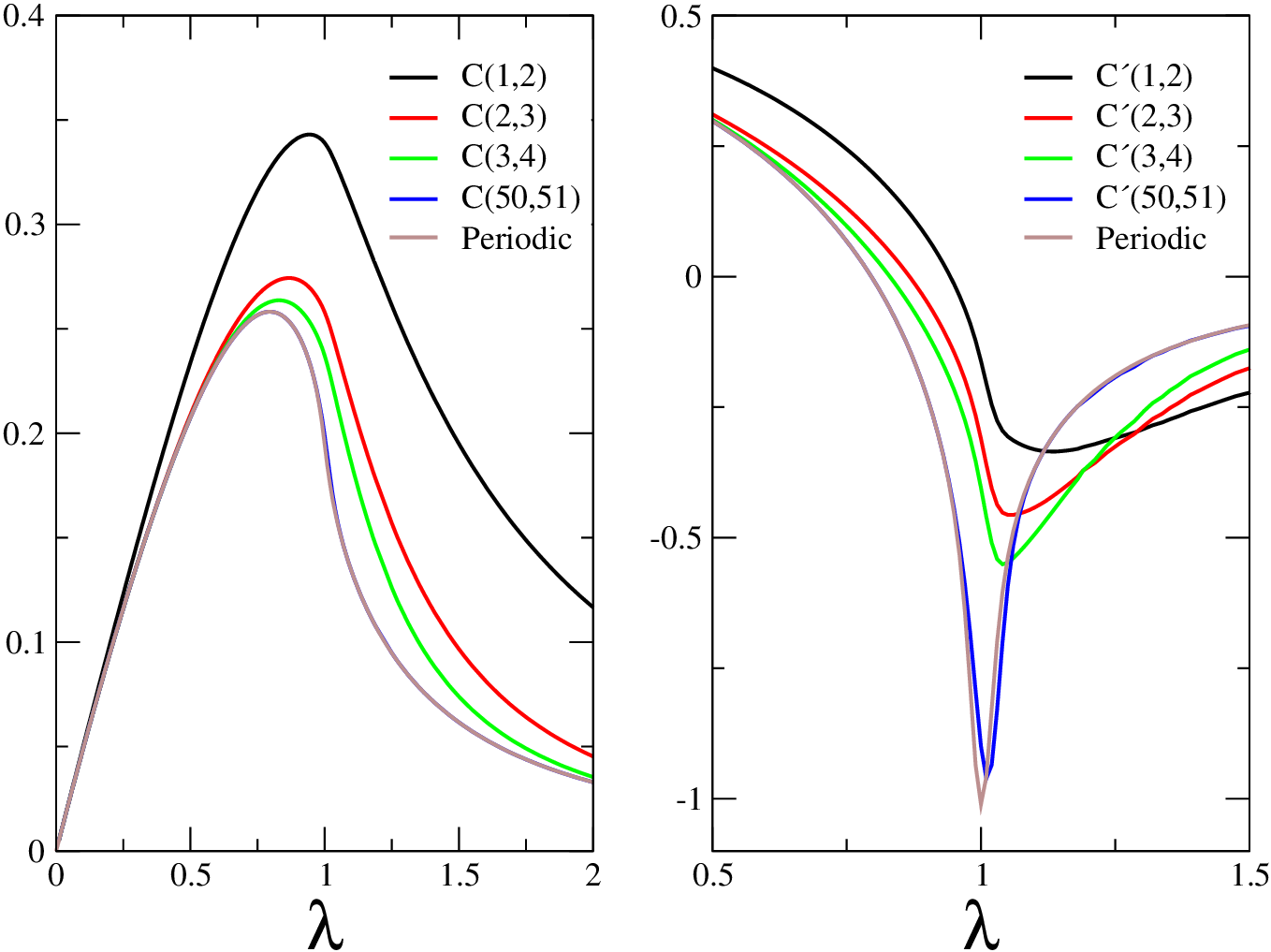}
    \caption{Left hand side: The nearest neighbor concurrence
    of the open boundary Ising model for different locations relative
    to the end as function of $\lambda$ for $N=101$. Right hand side:
    The derivative of the concurrence with respect to $\lambda$.}
    \label{Boundary}
\end{center}
\end{figure}
\begin{figure}[t]
  \begin{center}
    \includegraphics*[width=3in,angle=0]{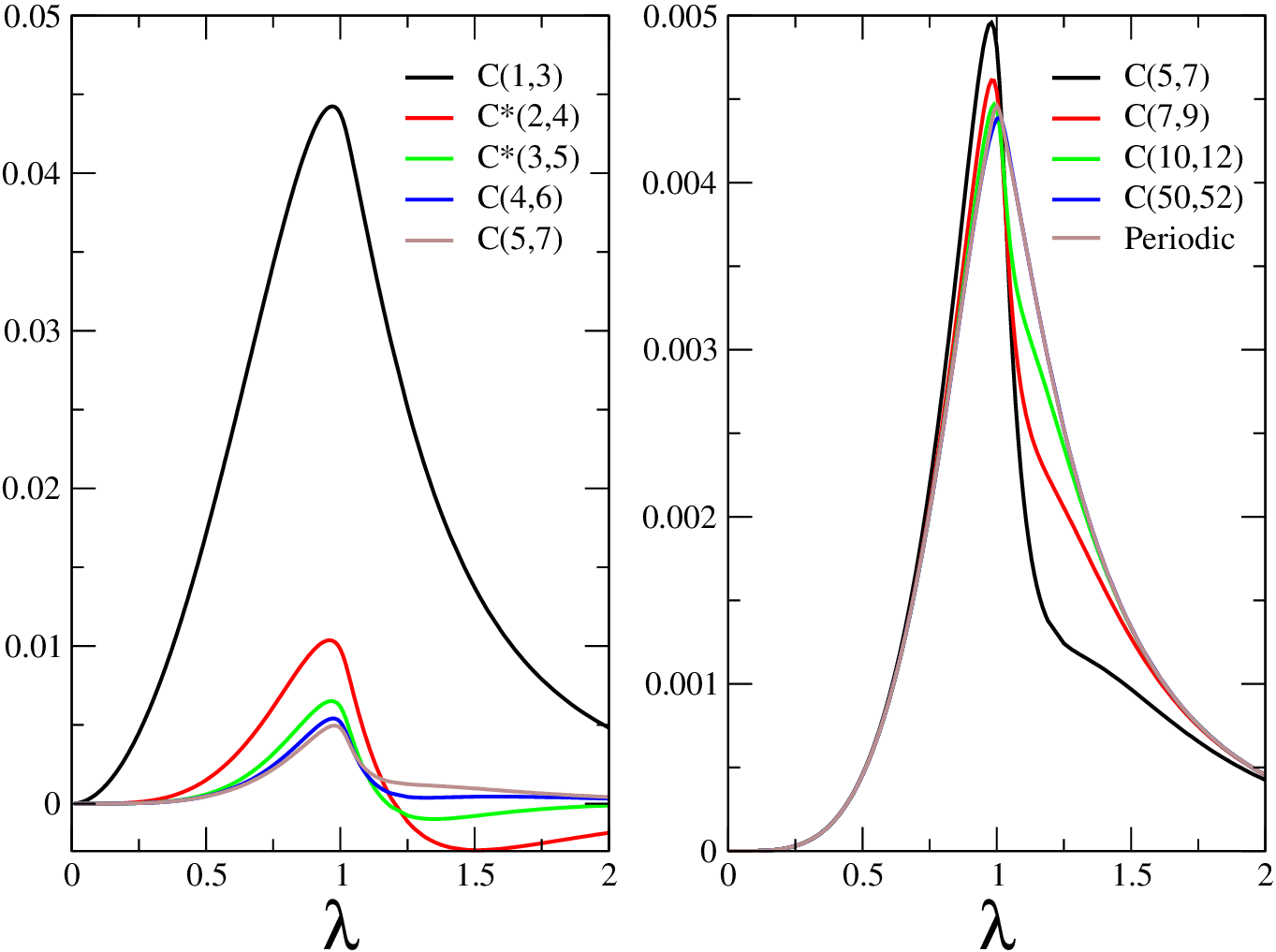}
    \caption{The next-nearest neighbor concurrence of the open
    boundary Ising model for different locations relative to the
    end of the chain as function of $\lambda$ for $N=101$.}
    \label{NextNearest}
\end{center}
\end{figure}
We first consider the nearest neighbor concurrence of the Ising
chain with open boundaries ($\kappa=0$) for a fixed number of
sites $N=101$ as parameter of $\lambda$, but for various positions
relative to the end of the chain. The results are displayed on the
left hand side of Fig. \ref{Boundary}. As expected, the
concurrence of the periodic model is approached as one moves
inside the chain and the difference between ${\cal C}(50,51)$ and
${\cal C}(i,i+1)$ of the periodic system is hardly seen.
Nevertheless, the derivative of the concurrence with respect to
the coupling parameter $\lambda$, ${\cal C}'\equiv d{\cal
C}/d\lambda$, still shows appreciable differences for
$\lambda\approx1$ (right hand side of Fig. \ref{Boundary}).

We also investigated the scaling behavior of the minimum of ${\cal
C}'(1,2)$, $\lambda_{min}$, for different systems sizes up to
$N=231$. We did not find finite-size scaling behavior for the
position of the minimum as is the case for the translationally
invariant model\cite{Oetal02}. The curve of ${\cal C}'(1,2)$,
shown on the right hand side of Fig. \ref{Boundary}, is thus
already close to the curve for $N\to\infty$ with a broad minimum
around $\lambda_{min}\approx1.1$.

The absence of finite-size scaling of the concurrence is also
manifested in the case of the next-nearest neighbor concurrence
for different system sizes $N$. Whereas for the periodic system
the maximum of ${\cal C}(i,i+2)$ decreases monotonically for
$N\to\infty$,\cite{Oetal02} there is practically no change of
${\cal C}(1,3)$ of the open chain for $N\gapp51$.

In Fig. \ref{NextNearest}, the generalized next-nearest neighbor
concurrence ${\cal C}^*(i,i+2)$ of the open boundary Ising model
is shown for different locations relative to the end of the chain
as function of $\lambda$ for $N=101$. On the left hand side of
Fig. \ref{NextNearest}, results are shown for sites close to the
end of the chain. Notice that the generalized concurrence becomes
negative for $i=2,3$ for $\lambda>1$ which is not related to the
quantum phase transition. The crossover of the boundary behavior
to the bulk behavior is thus discontinuous. On the right hand side
of Fig. \ref{NextNearest}, the next-nearest neighbor concurrence
approaches the result of the system with periodic boundary
conditions as one moves inside the chain.

\subsubsection{Generalized boundary conditions}
\begin{figure}[t]
  \begin{center}
    \includegraphics*[width=3in,angle=0]{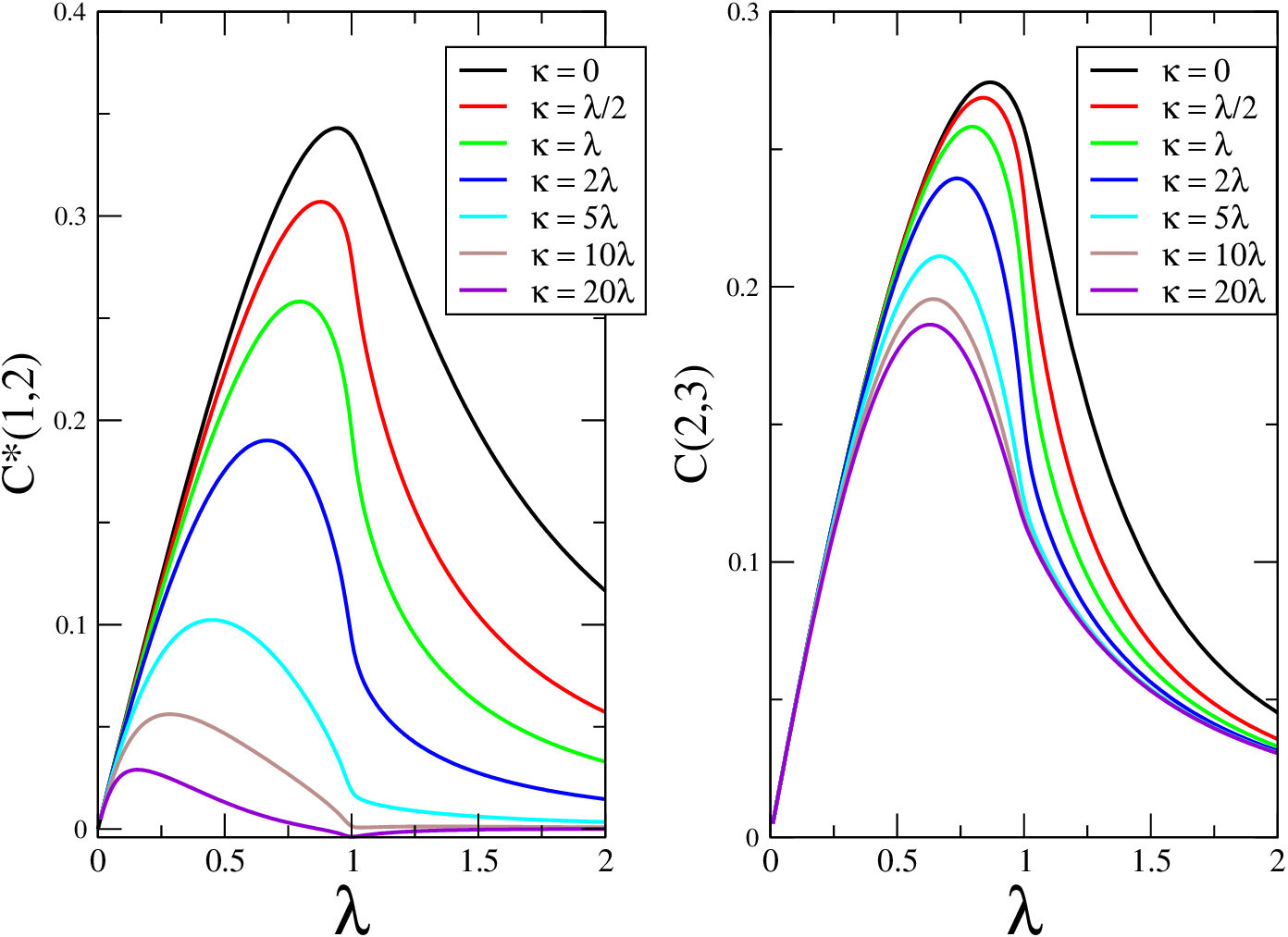}
    \caption{Left hand side: The (generalized) nearest neighbor concurrence of
    the closed Ising chain for various coupling strengths $\kappa$ as function
    of $\lambda$. Left hand side: ${\cal C}^*(1,2)$. Right hand side: ${\cal C}(2,3)$.}
    \label{Kopplung}
\end{center}
\end{figure}
\begin{figure}[t]
  \begin{center}
    \includegraphics*[width=3in,angle=0]{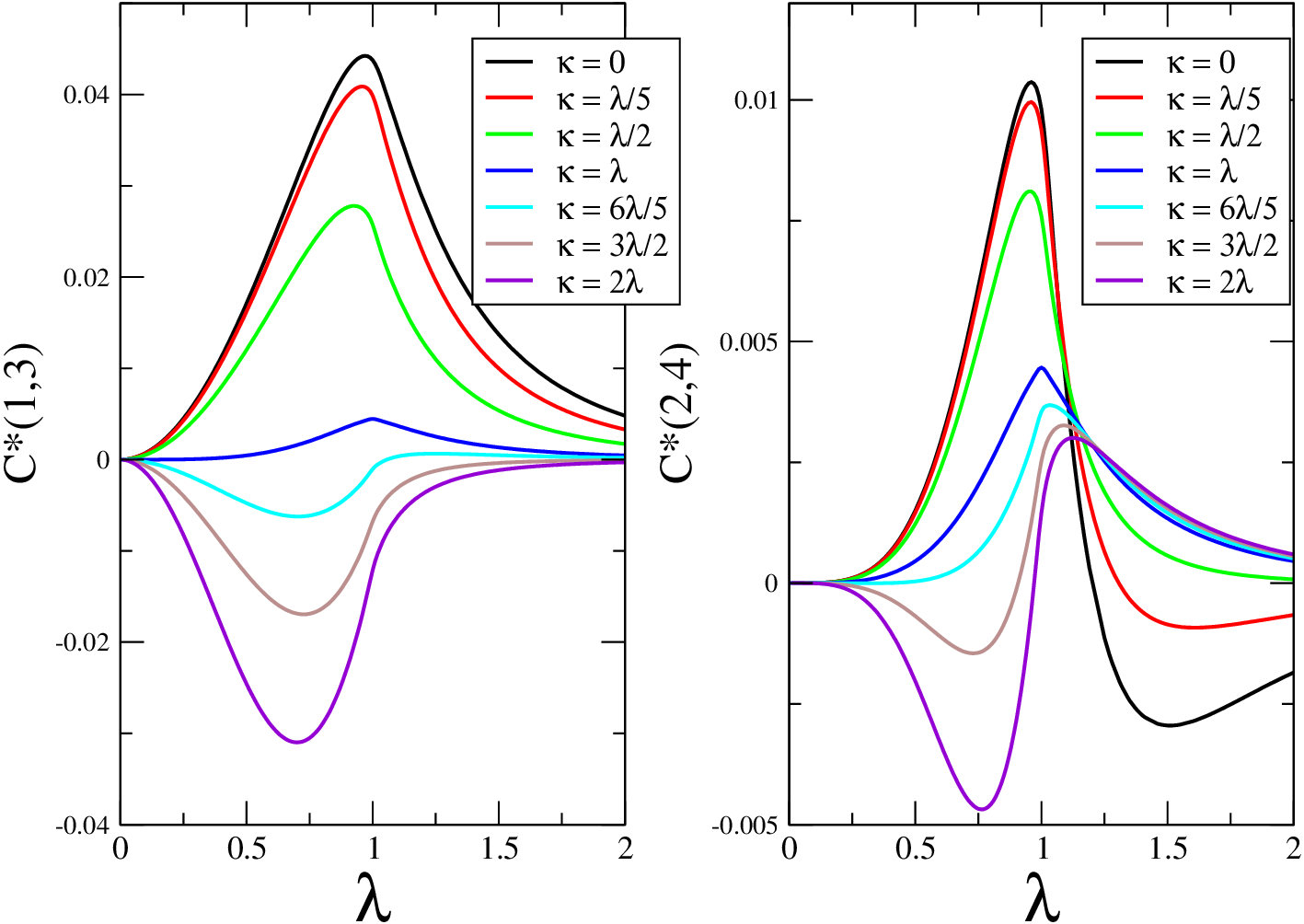}
    \caption{The generalized next-nearest neighbor concurrence of the closed Ising chain for
    various coupling strengths $\kappa$ as function of $\lambda$. Left hand side:
    ${\cal C}^*(1,3)$. Right hand side: ${\cal C}^*(2,4)$.}
    \label{Kopplung2}
\end{center}
\end{figure}
We now discuss the concurrence for the generalized boundary
conditions, introducing the parameter $\kappa$. On the left hand
side of Fig. \ref{Kopplung}, the generalized concurrence of the
first two spins ${\cal C}^*(1,2)$ is shown as function of
$\lambda$ for various coupling strengths $\kappa=0,..,20\lambda$
and $N=101$. For increasing $\kappa>0$, the curves indicate
stronger non-analyticity at $\lambda\approx1$. For $\kappa\gapp
20\lambda$, the generalized concurrence becomes negative around
$\lambda=1$ and is ''significantly'' positive only in the quantum
limit of a strong transverse field ($\lambda<1$). A similar
behavior of the concurrence is also found in the case of finite
temperatures.\cite{ON02}

On the right hand side of Fig. \ref{Kopplung}, the concurrence of
the second two spins ${\cal C}(2,3)$ is shown. All curves display
similar behavior. There is thus a rapid crossover from the
boundary to the bulk-regime and the concurrence of periodic
boundary conditions is approached for all $\kappa$ as one moves
further inside the chain.

To close, we discuss the next-nearest neighbor concurrence ${\cal
C}(i,i+2)$ for various values of $\kappa$ and $N=101$. On the left
hand side of Fig. \ref{Kopplung2}, the generalized concurrence of
the first and the third spin, ${\cal C}^*(1,3)$, is shown. For
$\kappa\lapp1$, ${\cal C}^*(1,3)$ is positive for all $\lambda$.
For $\kappa\gapp1$, ${\cal C}^*(1,3)$ first becomes negative for
$\lambda<1$. For $\kappa\gapp1.5$ ${\cal C}^*(1,3)$ is negative
for all $\lambda$. On the right hand side of Fig. \ref{Kopplung2},
the generalized concurrence of the second and the forth spin,
${\cal C}^*(2,4)$, is shown. For $\kappa\lapp\lambda/2$, the
${\cal C}^*(2,4)$ is negative for $\lambda\gapp1$. For
$\kappa\gapp3\lambda/2$, the ${\cal C}^*(2,4)$ is negative for
$\lambda\lapp1$. Nevertheless, the maximum value is close to
$\lambda=1$ for all cases.

We finally note that the third neighbor concurrence remains zero for all $\lambda$ and all $\kappa$.

\section{Boundary Phase Transition}
\subsection{The model.}
In order to observe critical behavior of the concurrence at the
boundary, we thus have to consider a different model, i.e., we
have to introduce an isotropic coupling from spin $N$ to spin $1$.
This will introduce an interaction term which contains four
fermionic operators and a simple solution is thus not possible
anymore.

The model with isotropic coupling is similar to the model
considered by Garst {\it et. al.}, who discuss two Ising-coupled
Kondo impurities\cite{Getal03} (see also\cite{VBH02}). We will
consider the model studied in\cite{Getal03}. It describes to Kondo
impurities, attached to two different electronic reservoirs, which
interact among themselves through an Ising term. We can write the
Hamiltonian as:
\begin{eqnarray}
{\cal H} &= &{\cal H}_{K1} + {\cal H}_{K2} + I S_{z1} S_{z2}
\nonumber \\
{\cal H}_{Ki} &= &\sum_k  \epsilon_{k , \mu} c^{\dag}_{k,\mu ,i}
c_{k, \mu , i} + J \sum_{k , k' , \mu , \nu} c_{k , \mu , i}
\vec{\sigma}_{\mu , \nu} c_{k', \nu , i} \vec{S}_i \nonumber \\ &
& \label{hamil}
\end{eqnarray}
We consider the entanglement of the two spins, by writing the $4
\times 4$ reduced density matrix in terms of the values of
$S_{z1}$ and $S_{z2}$.

The system described by Eq.(\ref{hamil}) undergoes a
Kosterlitz-Thouless transition between a phase with a doubly
degenerate ground state and a phase with a non degenerate ground
state. This transition is equivalent to that in the dissipative
two level system\cite{Letal87,W99} as function of the strength of
the dissipation. We define the dissipative two level system as:
\begin{equation}
{\cal H}_{TLS} = \Delta \sigma_x + \sum_k | k | b^{\dag}_k b_k +
\lambda \sigma_z \sum_k \sqrt{k} ( b^{\dag}_k + b_k ) \label{TLS}
\end{equation}
The strength of the dissipation can be characterized by a
dimensionless parameter, $\alpha \propto \lambda^2$, and the model
undergoes a transition for $\tilde{\Delta} = \delta / \omega_c \ll
1$, where $\omega_c$ is the cutoff, and $\alpha = 1$. The Kondo
model can be mapped onto this model\cite{GHM95} by taking $\Delta
\propto \tilde{J_\perp}$ and $1 - \alpha \propto \tilde{J_z}$.

To understand the equivalence between these two models, it is best
to to consider the limit $I/J \gg 1$ (the transition takes place
fr all values of this ratio). Let us suppose that $I > 0$ so that
the Ising coupling is antiferromagnetic. The Hilbert space of the
two impurities has four states. The combinations $| \uparrow
\uparrow \rangle$ and $| \downarrow  \downarrow \rangle$ are
almost decoupled from the low energy states, and the transition
can be analyzed by considering only the $| \uparrow \downarrow
\rangle $ and $ | \downarrow \uparrow \rangle$ combinations. Thus,
we obtain an effective two state system. The transition is driven
by the spin flip processes described by the Kondo terms. These
processes involve two simultaneous spin flips in the two
reservoirs. Hence, the operator which induces these spin flips
leads to the correspondence $\tilde{\Delta} \leftrightarrow 
J_\perp^2 / ( I \omega_c )$. The scaling dimension of this term,
in the Renormalization Group sense, is reduced with respect to the
ordinary Kondo Hamiltonian, as two electron-hole pairs must be
created. This implies the equivalence $2 - \alpha \leftrightarrow
\tilde{J_z}$. Hence, the transition, which for the ordinary Kondo
system takes place when changing the sign of $J_z$ now requires a
finite value of $J_z$.
\subsection{Calculation of the concurrence.}
The $4 \times 4$ reduced density matrix can be decomposed into a
$2 \times 2$ box involving the states $| \uparrow \downarrow
\rangle$ and $| \downarrow \uparrow \rangle$, which contains the
matrix elements which are affected by the transition, and the
remaining elements involving $| \uparrow \uparrow \rangle$ and $|
\downarrow \downarrow \rangle$ which are small, and are not
modified significantly by the transition. Neglecting these
couplings, we find that two of the four eigenvalues of the density
matrix are zero. The other two are determined by the matrix:
\begin{equation}
\tilde{\rho} \equiv \left( \begin{array}{cc} \frac{1}{2} + \langle
\sigma_z \rangle &\langle \sigma_x \rangle \\ \langle \sigma_x
\rangle &\frac{1}{2} - \langle \sigma_z \rangle \end{array}
\right) \label{matrix}
\end{equation}
where the operator $\tilde{\sigma}$ is defined using the standard
notation of the dissipative two level system, Eq.(\ref{TLS}). The
entanglement can be written as:
\begin{equation}
{\cal C} = \sqrt{ \langle \sigma_z \rangle^2 + \langle \sigma_x
\rangle^2} \label{entanglement_TLS}
\end{equation}
The value of $\langle \sigma_z \rangle$ is the order parameter of
the transition. The value of $\langle \sigma_x \rangle$, at zero
temperature, can be calculated from:
\begin{equation}
\langle \sigma_x \rangle = \frac{\partial E}{\partial \Delta}
\label{der_ener}
\end{equation}
where $E$ is the energy of the ground state. Using renormalization
group arguments, it can be written as:
\begin{equation}
E = \left\{ \begin{array}{lr} \frac{C}{1 - 2
\alpha} \left[ \Delta \left( \frac{\Delta}{\omega_c}
\right)^{\frac{\alpha}{1 - \alpha}} - \frac{\Delta^2}{\omega_c}
\right] &0 < \alpha < \frac{1}{2} \\ 2 C
\frac{\Delta^2}{\omega_c} \log \left( \frac{\omega_c}{\Delta}
\right) &\alpha = \frac{1}{2}
\\ \frac{C}{2 \alpha - 1} \left[ \frac{\Delta^2}{\omega_c} - \Delta
\left( \frac{\Delta}{\omega_c} \right)^{\frac{\alpha}{1 - \alpha}}
\right] &\frac{1}{2} < \alpha < 1 \\ C \left(
\frac{\Delta^2}{\omega_c} - C' \omega_c e^{- \frac{C''
\omega_c}{\Delta}} \right) &\alpha \sim 1 \\ C
\frac{\Delta^2}{\omega_c} & \alpha > 1
\end{array} \right. \label{ener_TLS}
\end{equation}
where $C , C'$ and $C''$ are numerical constants.

If the density matrix is calculated in the absence of a symmetry
breaking field, $\langle \sigma_z \rangle = 0$ even in the ordered
phase. Then, from Eq.(\ref{entanglement_TLS}), the concurrence is
given by ${\cal C} = |\langle \sigma_x \rangle|$, which is
completely determined using Eqs.(\ref{der_ener}) and
(\ref{ener_TLS}). In the limit $\Delta / \omega_c \ll 1$ the
interaction with the environment strongly suppresses the
entanglement. We expect unusual behavior of the concurrence for
$\alpha = 1 / 2$ and $\alpha = 1$. The point $\alpha = 1/2$ marks
the loss of coherent oscillations between the two
states\cite{G85,note2}, although the ground state remains non
degenerate. Following the analysis in\cite{Oetal02}, we analyze
the behavior of $\partial {\cal C} /
\partial \alpha$, as $\alpha$ is the parameter which determines
the position of the critical point. The strongest divergence of
this quantity occurs for $\alpha = 1 / 2$, where:
\begin{equation}
\left. \frac{\partial {\cal C}}{\partial \alpha} \right|_{\alpha =
1/2} \sim \frac{\Delta}{\omega_c} \log \left(
\frac{\omega_c}{\Delta} \right)
\end{equation}
On the other hand, near $\alpha = 1$ the value of $\partial {\cal
C} / \partial \alpha$ is continuous, as the influence of the
critical point has a functional dependence, when $\alpha
\rightarrow \alpha_c$, of the type $e^{- c / ( \alpha_c - \alpha
)}$. This is the standard behavior at a Kosterlitz-Thouless phase
transition. This result suggest that the entanglement is more
closely related to the presence of coherence between the two
qubits than with the phase transition. The transition takes place
well after the coherent oscillations between the $| \uparrow
\downarrow \rangle$ and $| \downarrow \uparrow \rangle$ states are
completely suppressed.

\section{Summary}
We have first calculated the entanglement between qubits at the
boundary of a spin chain, whose parameters are tuned to be near a
quantum critical point. The calculations show a behavior which
differs significantly from the that inside the bulk of the chain.
Although the spins are part of the critical chain, we find no
signs of the scaling behavior which can be found in the bulk. We
use the same approach as done previously for the
bulk\cite{Oetal02,ON02}, although it should be noted that the
existence of a finite order parameter in the ordered phase will
change these results if the calculations were performed in the
presence of an infinitesimal applied field.

We have also considered the entanglement between two spins coupled
to a dissipative environment and which undergo a local quantum
phase transition. The system which we have studied belongs to the
generic class of systems with a Kosterlitz-Thouless transition at
zero temperature, like the Kondo model or the dissipative two
level system. The most remarkable feature of our results is that
the entanglement properties show a pronounced change at the
parameter values where the coherent quantum oscillations between
the qubits are lost, and not at the location of the proper phase
transition, where the ground state becomes degenerate.  At this
point, however, the interaction with the environment has rendered
the dynamics of the qubits extremely incoherent.

\section{Acknowledgments}
We are grateful to J. I. Latorre and to M. A. Mart{\'\i}n-Delgado
for a critical reading of the manuscript. T.S. acknowledges
support from the EU-RTN under ``HPRN-CT-2000-00144''. Funding from
MCyT (Spain) through grant MAT2002-0495-C02-01 is also
acknowledged.
\bibliography{Ising3}

\begin{thebibliography}{22}
\bibitem[{bqc()}]{bqcp}
\bibinfo{note}{These models can be mapped onto one dimensional systems with
  long range interactions. They can only undergo a phase transition at zero
  temperature.}
\bibitem[{not()}]{note2}
\bibinfo{note}{It is interesting to note that the effective value of $\alpha$
  for the Ising model considered in the previous section is $\alpha = 1 / 2$,
  as, at criticality, $\langle \sigma_x ( t ) \sigma_x ( t' ) \rangle \sim ( t
  - t' )^{-1}$.}
\end{thebibliography}
\end{document}